\documentclass[a4paper,fleqn,usenatbib]{mnras}

\usepackage[T1]{fontenc}
\usepackage{ae,aecompl}

\usepackage{graphicx}	% Including figure files
\usepackage{amsmath}	% Advanced maths commands
\usepackage{amssymb}	% Extra maths symbols

\newcommand{\bea}{\begin{eqnarray}}
\newcommand{\eea}{\end{eqnarray}}
\newcommand{\mrm}{\mathrm}

\title[Optimising primordial non-Gaussianity measurements from galaxy surveys]{Optimising primordial non-Gaussianity measurements from galaxy surveys}

\author[E.-M. Mueller]{
Eva-Maria Mueller,$^{1}$\thanks{E-mail: eva-maria.mueller@port.ac.uk}
Will J. Percival,$^{1}$
Rossana Ruggeri$^{1}$
\\
$^{1}$Institute of Cosmology \& Gravitation, University of Portsmouth, Dennis Sciama Building, Portsmouth, PO1 3FX, UK\\
}

\date{Accepted XXX. Received YYY; in original form ZZZ}

\pubyear{2017}

\begin{document}
\label{firstpage}
\pagerange{\pageref{firstpage}--\pageref{lastpage}}
\maketitle

%%%%%%%%%%%%%%%%%%% ABSTRACT %%%%%%%%%%%%%%%%%%%
\begin{abstract}
Galaxy clustering data from current and upcoming large scale structure surveys can provide strong constraints on primordial non-Gaussianity through the scale-dependent halo bias. To fully exploit the information from galaxy surveys, optimal analysis methods need to be developed and applied to the data. Since the halo bias is sensitive to local non-Gaussianity predominately at large scales, the volume of a given survey is crucial. Consequently, for such analyses we do not want to split into redshift bins, which would lead to information loss due to edge effects, but instead analyse the full sample. We present an optimal technique to directly constrain local non-Gaussianity parametrised by $f_\mrm{NL}^\mrm{loc}$, from galaxy clustering by applying redshift weights to the galaxies.
We derive a set of weights to optimally measure the amplitude of local non-Gaussianity, $f_\mrm{NL}^\mrm{loc}$, discuss the redshift weighted power spectrum estimators, outline the implementation procedure and test our weighting scheme against Lognormal catalogs for two different surveys: the quasar sample of the Extended Baryon Oscillation Spectroscopic Survey (eBOSS) and the emission line galaxy sample of the Dark Energy Spectroscopic Instrument (DESI) survey. We find an improvement of 30 percent for eBOSS and 6 percent for DESI compared to the standard Feldman, Kaiser \& Peacock weights, although these predictions are sensitive to the bias model assumed.

\end{abstract}


%%%%%%%%%%%%%%%%%%%%%%%%%%%%%%%%%%%%%%%%%%%%%%%%%%

%%%%%%%%%%%%%%%%% BODY OF PAPER %%%%%%%%%%%%%%%%%%

%%%%%%%%%%%%%%%%%%% INTRODUCTION %%%%%%%%%%%%%%%%%%%

\section{Introduction}

Primordial non-Gaussianity (PNG) is one of the most promising probes to distinguish between different models of inflation, a theory to describe an era of exponential expansion of the very early universe that was first introduced to solve problems within the Big Bang model. Inflation can solve the horizon problem as well as the flatness problem, and can also explain the origin of structure formation through the creation of initial fluctuations. Currently, the best constraints on PNG are provided by measurements of the cosmic microwave background (CMB) with the Planck satellite \citep{2016A&A...594A..17P}.  

Even though current constraints from large scale structure (LSS) data (i.g. \cite{2013MNRAS.428.1116R}) are weaker than the CMB results, future galaxy surveys have the potential to significantly improve upon these limits (see e.g. \cite{2008ApJ...684L...1C, 2011MNRAS.414.1545F, 2012MNRAS.422.2854G, 2014arXiv1412.3854D, 2015MNRAS.448.1035C, 2015PhRvD..91d3506F, 2015JCAP...03..019B, 2012JCAP...12..034B,2012MNRAS.422.2854G, 2015JCAP...08..034R}) by constraining the scale dependent halo bias induced by PNG \citep{2008PhRvD..77l3514D, 2008ApJ...677L..77M,2008JCAP...08..031S, 2010CQGra..27l4011D}. Upcoming spectroscopic surveys such as the extended Baryon acoustic Oscillation Spectroscopic Survey (eBOSS) \citep{2016MNRAS.457.2377Z}, the Euclid mission \citep{2013LRR....16....6A}, as well as the Dark Energy Spectroscopic Instrument (DESI) \citep{2014JCAP...05..023F} survey are expected to constrain the amplitude of local non-Gaussianity, $f_\mrm{NL}^\mrm{loc}$, around a few (from here on we will drop the subscript 'loc' for simplicity); however, to achieve that level of accuracy, analysing techniques need to be optimised to fully exploit the LSS information. Indeed, most galaxy redshift survey analyses fall short of their expected results predicted using Fisher Matrix techniques. 

%Redshift weighting, a recent analysing strategy that relies on assigning weights to the galaxies in a particular sample, taking the redshift evolution of the underlying physical theory into account instead of splitting up the sample in redshift shells, has the potential to notably improve cosmological constraints from LSS surveys. 
It was recently realised \citep{2015MNRAS.451..236Z} that that some of the missing signal is lost because analyses are generally performed after splitting a galaxy sample into redshift shells. Instead, they proposed adopting an analysis strategy that relies on assigning weights to the galaxies over a broad redshift range, showing that this retains more information provided that the weights take the redshift evolution of the underlying physical theory into account. This has the potential to notably improve cosmological constraints from LSS surveys.

 \cite{2015MNRAS.451..236Z} focussed on optimising LSS surveys for BAO measurements, and their method was shown to work using mock catalogs in \cite{2016MNRAS.461.2867Z}. In subsequent work, redshift weights were derived to constrain modified gravity through Redshift Space Distortions (RSD) in \cite{2017MNRAS.464.2698R}. These weights can be interpreted as a natural extension of the Feldman, Kaiser \& Peacock (FKP) weights \citep{1994ApJ...426...23F}, that balance galaxies according to their number densities, for the case that the cosmological observables of interest evolve with time. If the underlying physical theory is independent of redshift then the weights reduce to the standard FKP weights. In the future, multiple galaxy surveys will cover a large redshift range, $0<z<3$, making the redshift weighting technique particularly efficient as well as necessary to avoid information loss due to edge effects and disjoint bins. Furthermore, the computational time can be reduced significantly since the redshift weighting technique only requires a single analysis instead of measuring each redshift bin separately. Redshift weighting also removes the need to define an effective redshift of a survey by providing measurements with known variation over the redshift range.

In this paper, we derive and assess the redshift weights for optimising LSS surveys for local $f_\mrm{NL}$ measurements. Avoiding redshift binning is particularly relevant for non-Gaussianity measurements since the effect of the scale dependent bias dominants on very large scales. Breaking the survey into redshift bins (for example, considering the clustering in bins of width $\Delta z=0.1$), removes large-scale clustering signal. For the correlation function it is clear that such binning removes pairs of galaxies, where galaxies lie in different bins. For the power spectrum, the binning introduces a window function, correlating large-scale modes, and decreasing the effective number of modes.

The paper is organised as follows: In Section \ref{sec:physical_model} we summarise modelling of the power spectrum as well as the observable effects of non-Gaussianity on the power spectrum. We introduce the concept of redshift weighting in Section~\ref{sec:weights} and derive the optimal weights for $f_\mrm{NL}$ measurements in Section~\ref{sec:weights_fNL}. We outline the procedure of how to apply the weights to the data in Section~\ref{sec:procedure}. In Section~\ref{sec:P0weighted} we discuss the modelling of the redshift weighted power spectrum and in Section~\ref{sec:Testingweights} we estimate the improvement of using $f_\mrm{NL}$ weights compared to FKP weights by simulating the redshift weighted power spectrum estimators using Lognormal catalogs. Finally, Section~\ref{sec:conclusion} contains the discussion of our results and conclusions. 

%%%%%%%%%%%%%%%%%%% PHYSICAL MODEL %%%%%%%%%%%%%%%%%%%

%%%%%%%%%%%%%%%%%%%%%%%%%%%%%%%%
\section{Physical Model}
\label{sec:physical_model}
%%%%%%%%%%%%%%%%%%%%%%%%%%%%%%%%
In this Section we provide a brief summary of the scale dependent halo bias induced by non-Gaussianity. In the framework of local non-Gaussianity, i.e. a type of non-Gaussianity that only depends on the local value of the potential, the primordial potential can be parametrised as \citep{2001PhRvD..63f3002K,1994ApJ...430..447G}
\bea
\Phi = \phi + f_\mrm{NL} (\phi^2 - \langle \phi^2 \rangle)
\eea
where $\phi$ is a gaussian random field and $f_\mrm{NL}$ describes the amplitude of the quadratic correction to the potential. The potential can then be related to the density field via $\delta(k)=\alpha(k) \Phi(k)$, with
\bea
\alpha(k) = \frac{2 k^2 T(k) D(z)}{3\Omega_m} \frac{c^2}{H_0^2} \frac{g(0)}{g(\infty)} \label{eq:alpha}
\eea
with the transfer function $T(k)$, the linear growth factor $D(z)$ normalised to be unity at $z=0$, the matter density today $\Omega_m$, the speed of light c and the Hubble parameter today $H_0$. The factor $g(\infty)/g(0)$, with $g(z)=(1+z)D(z)$, arises due to our normalisation of $D(z)$ and can be omitted if D(z) is normalised to equal the scale factor during the matter dominated era. Here we are using the CMB convention for $f_\mrm{NL}$ assuming $\Phi$ is the primordial potential. Note that some authors have previously adopted a "LSS convention" that assumes $\Phi$ is extrapolated to $z=0$, with $f_\mrm{NL}^\mrm{LSS}=g(\infty)/g(0) f_\mrm{NL}^\mrm{CMB} \approx 1.3  \ f_\mrm{NL}^\mrm{CMB}$. We do not do this as we consider it unnecessary and potentially confusing.
 
 The scale dependent halo bias $\Delta b(k)$ in the local Ansatz is then given by \citep{2008PhRvD..77l3514D,2008JCAP...08..031S}
\bea
\Delta b(k)= 2 (b-p)f_\mrm{NL} \frac{\delta_\mrm{crit}}{\alpha(k)} \label{eq:db}
\eea
where $\delta_\mrm{crit}=1.686$ and $1<p<1.6$ depending on the type of tracer. Here we follow \cite{2008JCAP...08..031S} assuming $p=1$ for luminous red galaxies (LRGs) and emission line galaxies (ELGs) and $p=1.6$ for quasars.

The total bias, including local non-Gaussianity is then
$
b_\mrm{tot} = b + \Delta b(k).
$

In the limit of the plane parallel approximation, the linear matter power spectrum $P$ in redshift space is \citep{1987MNRAS.227....1K}
\bea
P(k,\mu) = \left(b_\mrm{tot} + f \mu^2\right)^2 P_M(k) \label{eq:ps}
\eea
where $f$ is the linear growth rate, $\mu$ is the cosine of the angle between the wavevector k and the line of sight and $P_M(k)$ is the linear matter power spectrum. The effect of $f_\mrm{NL}$ is included in the definition of the total bias. From an observational point of view it is more convenient to consider the power spectrum multipoles defined as
\bea
P_l(k) = \frac{2l+1}{2}\int^{1}_{-1}d\mu P(k,\mu) \mathcal{L}_l(\mu)
\eea
where $\mathcal{L}_l(\mu)$ are the Legendre polynomials, instead of the linear power spectrum given by equation~(\ref{eq:ps}). Even though the power spectrum is fully defined by its first three moments at linear order, only the monopole
\bea
P_0(k)=\left(b_\mrm{tot}^2+\frac{2}{3} fb_\mrm{tot}+\frac{1}{5} f^2\right)P_M(k) \label{eq:P0}
\eea
as well as the quadrupole
\bea
P_2(k)= \left(\frac{4}{3}b_\mrm{tot}f+\frac{4}{7}f^2\right)P_M(k) \label{eq:P2}
\eea
depend on the bias. Therefore we focus our analysis to these multipoles.

%%%%%%%%%%%%%%%%%%% OPTIMAL WEIGHTS%%%%%%%%%%%%%%%%%%%

%%%%%%%%%%%%%%%%%%%%%%%%%%%%%%%%%%%
\section{Optimal weights}
%%%%%%%%%%%%%%%%%%%%%%%%%%%%%%%%%%%
%%%%%%%%%%%%%%%%%%%%%%%%%%%%%%%%%%%
\subsection{Redshift weights}
\label{sec:weights}
%%%%%%%%%%%%%%%%%%%%%%%%%%%%%%%%%%%
%########################FIG1##########################
\begin{figure*}
	\includegraphics[width=0.49\linewidth]{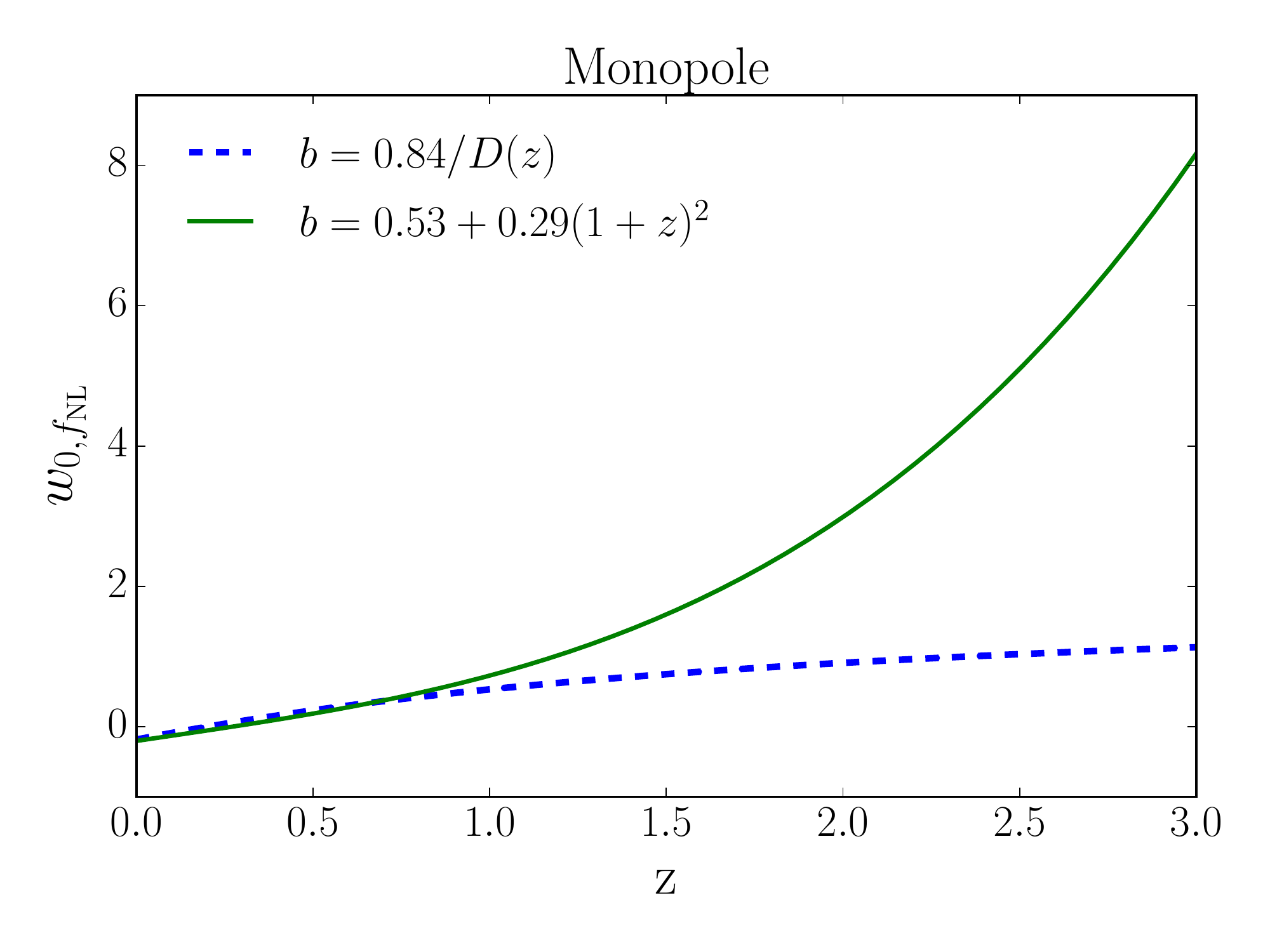}
	\includegraphics[width=0.49\linewidth]{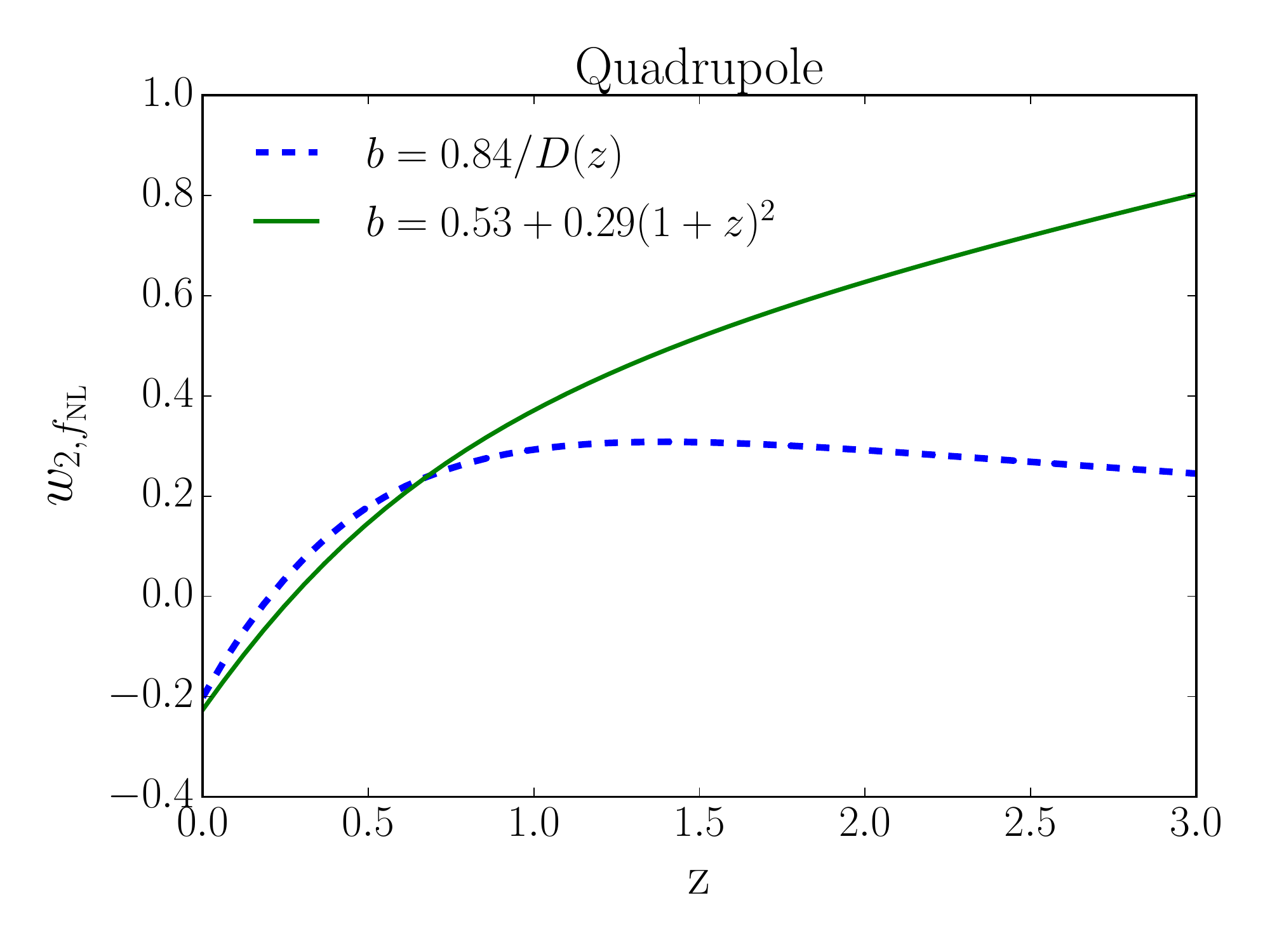}
  	\caption{Optimal redshift weights for local non-Gaussianity, $f_\mrm{NL}$, measurements as a function of redshift for the power spectrum monopole [left panel] and quadrupole [right panel]. Blue dashed lines assume a bias model $b(z)=0.53+0.29(1+z)^2$ and $p=1.6$ referring to quasars as tracer of the underlying matter density, while green lines are for $b(z)=0.84/D(z)$ and $p=1.0$, referring to ELGs. The assumptions on the fiducial value of $f_\mrm{NL}$ have an negligible effect on the weights. For these plots we assume a fiducial of $f_\mrm{NL}=0$  }
		\label{fig:wfNL}
\end{figure*}
%#######################################################
Following the procedure outlined in \cite{2017MNRAS.464.2698R} and \cite{2016MNRAS.461.2867Z} we can derive the optimal redshift weights by maximising the Fisher information matrix F defined as
\bea
F_{ij} \equiv \left \langle \frac{\partial^2 \mathcal{L}}{\partial \theta_i \partial \theta_j} \right \rangle, \\
\eea  
with the likelihood function $\mathcal{L}$ and the parameters $\theta_i$. Assuming a Gaussian likelihood, the fisher matrix for a single parameter of the weighted data set can be calculated as
\bea
F_{ii} =  \frac{1}{2} \left(\frac{ \bold{w}^T C_{,i} \bold{w}}{\bold{w}^TC\bold{w}} \right)^2 +  \frac{(\bold{w}^T\mu_{,i})^2}{\bold{w}^TC\bold{w}}\label{eq:Fij} 
\eea
with the covariance matrix $C$, the mean $\mu$, the weights $\bold{w}$ and the index ${,i}$ denoting the partial derivative $\partial/\partial \theta_i$ \citep[see e.g.][]{1997ApJ...480...22T,1996ApJ...465...34V}.

The first term in equation (\ref{eq:Fij}) vanishes assuming the covariance matrix is known and independent of the cosmological parameters. The second term is maximised for 
\bea
\bold{w}^T=C^{-1} \mu_{,i}. \label{eq:weight}
\eea
Defining $w_i= \mu_{,i}^T$ as well as $d\mathcal{W} \equiv C^{-1}$ the weights for parameter $\theta_i$ can be written as
\bea
\bold{w}= w_i \  d\mathcal{W}.
\eea
The factor $d\mathcal{W}$ takes the statistical uncertainty of the observable into consideration whereas $w_i$ factors in the redshift evolution of the theoretical model. 
The normalisation of the weights is arbitrary and does not affect the cosmological constraints. We choose the normalisation, 
\bea
N_i=\int w_i \ d\mathcal{W} \label{eq:norm}
\eea
leading to normalised weights $\hat{ \bold{w}}$ defined as
\bea
\hat{ \bold{w}} \equiv \frac{1}{N_i} w_i \  d\mathcal{W}.
\eea
Note first, that the index i refers to the same cosmological parameter but does not imply Einstein summation and second, that our normalisation differs from the one in \cite{2016MNRAS.461.2867Z} by a factor of $w_i$.

%%%%%%%%%%%%%%%%%%%%%%%%%%%%%%%%%%%%
\subsection{Redshift weights for local non-Gaussianity}
\label{sec:weights_fNL}
%%%%%%%%%%%%%%%%%%%%%%%%%%%%%%%%%%%%
For the power spectrum P(k), the inverse covariance matrix in each redshift slice can be approximated by
\bea
d\mathcal{W} \equiv C^{-1}= \left(\frac{\bar{n}}{\bar{n}P+1}\right)^2 dV
\eea
depending on the galaxy density $\bar{n}$ as well as the survey volume $dV$. Since in this analysis we are interested in using measurements of the power spectrum monopole and quadrupole to constrain the non-Gaussianity parameter $f_\mrm{NL}$, the part of weights referring to the redshift evolution of $f_\mrm{NL}$ are given by
\bea
w_{l,f_\mrm{NL}} = \frac{\partial P_l}{\partial f_\mrm{NL}}.
\eea
The total weights
\bea
\bold{w} = w_{l,f_\mrm{NL}} d\mathcal{W}
\eea
are then a combination of the volume factor $d\mathcal{W}$ and the $f_\mrm{NL}$ weights. In the following we will use the term "redshift weights" to refer to 
$w_{l,f_\mrm{NL}}$ but one should keep in mind that the total weights also include the volume factor $d\mathcal{W}$.
For $w_i=1$ the weights reduce to the commonly used FKP weights. However, if one is interested in a theory that is more sensitive at high redshifts, for instance, more total weight will be given to galaxies at higher redshifts than in the case of FKP weights.  

Using equation~(\ref{eq:alpha}) and equation~(\ref{eq:db}) together with equation~(\ref{eq:P0}), the weight of the monopole reads as
\bea
w_{0,f_\mrm{NL}}&=& \left(2 b_\mrm{tot} +\frac{2}{3} f \right) \frac{\partial b_\mrm{tot}}{\partial f_\mrm{NL}} P_M(k,z) 
% &=& \left(2 b_\mrm{tot} +\frac{2}{3} f \right) 2(b-p)\frac{\delta_\mrm{crit}}{\alpha(k,z)} P_M(k,z) \
\eea
and furthermore assuming a fiducial value for $f_\mrm{NL,fid}=0$ simplifies to
\bea
w_{0,f_\mrm{NL}}&=& \left(2 b  +\frac{2}{3} f \right) 2(b-p)\frac{\delta_\mrm{crit}}{\alpha(k,z)} P_M(k,z).
\eea

Factoring out the explicit redshift dependency as $\alpha(k,z)=\alpha(k,z_0)D(z)$ and $P_M(k,z)=P_M(k,z_0) D(z)^2$ as well as normalising the weights according to equation~(\ref{eq:norm}), the normalised weights can be written independent of the wavevector $k$. Without the loss of generality, the weights can be redefined as
\bea
 \hat{w}_{0,f_\mrm{NL}} = \frac{1}{N_{0,f_\mrm{NL}}} w_{0,f_\mrm{NL}}
\eea
where 
\bea
w_{0,f_\mrm{NL}} = \left( b  +\frac{1}{3} f \right) (b-p) D(z) \\
N_{0,f_\mrm{NL}} = \int w_{0,f_\mrm{NL}}  \ d\mathcal{W}.
\eea

Similarly the quadrupole weight can be defined as
\bea
w_{2,f_\mrm{NL}}&=&\frac{4}{3} f (b-p) D(z) .
\eea

It should be emphasised the scale independence of these weights significantly simplifies their application (see Section~\ref{sec:procedure}).

Fig.~\ref{fig:wfNL} shows the weight for the monopole $w_{0,f_\mrm{NL}}$ [left panel] and the quadrupole $w_{2,f_\mrm{NL}}$ [right panel] as a function of redshift $z$ assuming a bias of $b(z)=0.53+0.29(1+z)^2$ (blue dashed lines) and $p=1.6$ as well as $b(z)=0.84/D(z)$ and $p=1$ (green lines), bias models previously proposed for eBOSS quasars \citep{2016MNRAS.457.2377Z} and DESI ELGs \citep{2016arXiv161100036D} respectively. The weights at low redshifts, $z<0.75$, are similar for both, but deviate for higher redshifts due to increasing differences in the bias models, with a strong high-redshift bias leading to larger weights at high redshifts. In general, the $f_\mrm{NL}$ weights are also larger for higher redshifts since the $f_\mrm{NL}$ model is also sensitive to the redshift evolution of the the growth rate.

\subsection{Implementation procedure}
\label{sec:procedure}

The implementation procedure was outlined in \cite{2015MNRAS.451..236Z} analysing the real space correlation function as well as in \cite{2017MNRAS.464.2698R} for the power spectrum in Fourier space. For completeness, we summarise some of the key-points here. The redshifts weights can be applied to the data and randoms as an extension of the usual FKP weighting scheme following the prescription of \cite{1994ApJ...426...23F}, 
\bea
w_\mrm{FKP} = \frac{1}{1+\bar{n}(z)P(k_0)}
\eea
where $\bar{n}(z)$ is the mean number density at the galaxies's redshift z, and $k_0$ is commonly assumed to be approximately the BAO scale. The redshift dependent weights are applied in the following way: In real space, each galaxy pair (or pair of randoms) is weighted by $w_{l,f_\mrm{NL}} $ as well as $w_\mrm{FKP}$
\bea
\widetilde{XY} = \sum_z w_{l,f_\mrm{NL}} w^2_\mrm{FKP}XY
\eea
where $\widetilde{XY}=\{DD,DR,RR\}$ refer to the data-data, data-random and random-random pairs of the sample. The standard \cite{1993ApJ...412...64L} estimator 
\bea
\xi_{l,f_\mrm{NL}} = \frac{\widetilde{DD}-2\widetilde{DR}+\widetilde{RR}}{RR}
\eea
can then be used to calculate the weighted correlation function, where $RR$ are the unweighted random-random pairs.

In Fourier space the procedure is similar. Each galaxy is weighted by a product of FKP weight and the $f_\mrm{NL}$ specific weights as derived in Section~\ref{sec:weights}
\bea
w = \sqrt{w_\mrm{FKP}  \times w_{l,f_\mrm{NL}}}.
\eea
Note, that even though we derived the weights within the framework of the power spectrum, following the assumption that the clustering evolves over larger scales than those being measured, we can approximate the weights applied to the galaxies as the root of the power spectrum weights $w_g= \sqrt{w_P}$.  %A possible negative sign of the weights could be absorbed into the normalisation which would require to divide the sample in two part. Yet in the case for $f_\mrm{NL}$, the weights are only negative for very small redshift  

%Following \cite{1994ApJ...426...23F} the standard estimator of the power spectrum $P(\bf{k})$ can be constructed as
%\bea
%\hat{P}(\bold{k}) = |F(\bold{k})|^2 - P_\mrm{shot}
%\eea
%with the shot noise $P_\mrm{shot}$ and the Fourier transform of the weighted galaxy fluctuation field, $F(\bold{r})$, defined as
%\bea
%F(\bold{r}) \equiv \frac{w(\bold{r})\left[n_g(\bold{r})-\alpha n_s(\bold{r}) \right]}{I^{1/2}}
%\eea
%where $n_g(\bold{r})$ and $n_s(\bold{r})$ are the observed number density of galaxies and the number density of a synthetic catalog which has number density $1/\alpha$ times that of the real catalogue, $w(\bold{r})$ is the weight applied to the galaxies and $I$ is a normalisation factor. This formalism can be generalised to include the redshift weighting by defining $F(\bold{r})$ with the additional weights included 
%\bea
%F(\bold{r}) &\longrightarrow& F(\bold{r},z) \\
%w(\bold{r}) &\longrightarrow& w(\bold{r},z)
%\eea
%with the galaxy weight given by the square root of the power spectrum weight,  $w(\bold{r},z)= \sqrt{w_P}$, with $w_P$ given by equation~\ref{eq:weight}.

%%%%%%%%%%%%%%%%%%%%%%%%%%%%%%%%%%%%%%%%
\subsection{Modelling the weighted power spectrum}
\label{sec:P0weighted}
%%%%%%%%%%%%%%%%%%%%%%%%%%%%%%%%%%%%%%%%
%########################FIG2##################################
\begin{figure*}
	\includegraphics[width=0.49\textwidth]{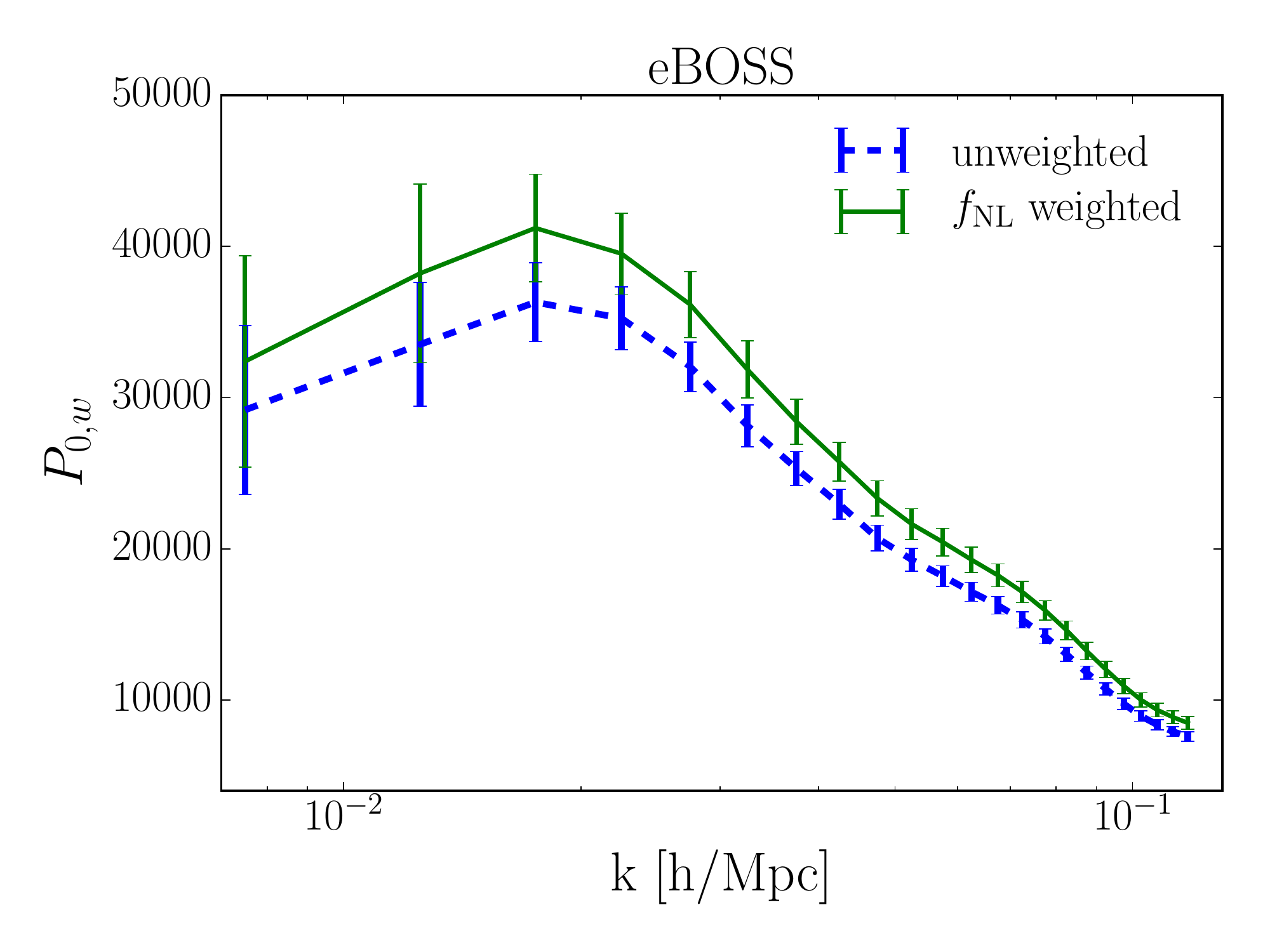}
	\includegraphics[width=0.49\textwidth]{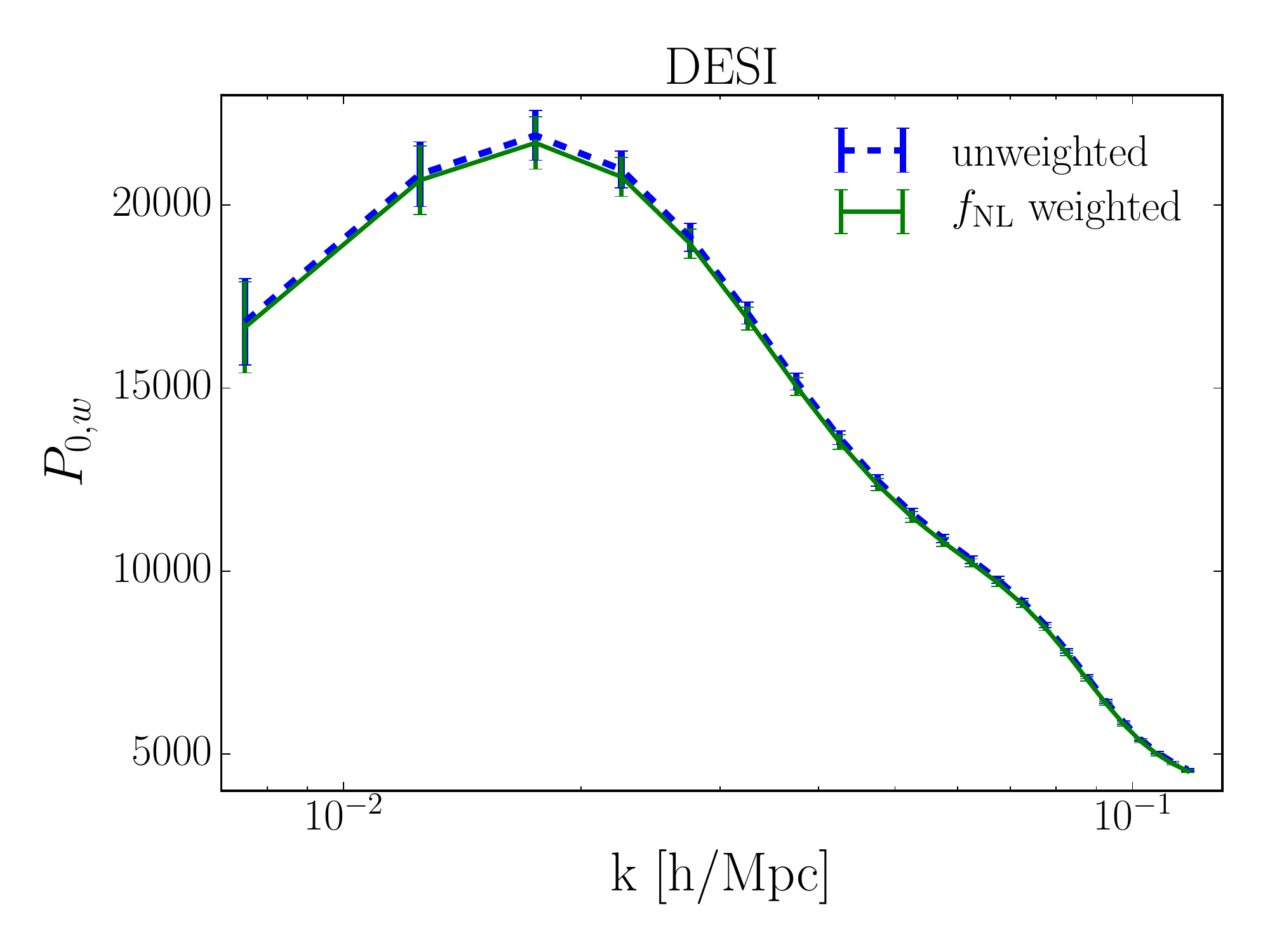}
  	\caption{Redshift weighted power spectrum monopole for eBOSS [left panel] and DESI [right panel]. Blue dashed lines correspond to the 'unweighted' (or FKP)-weighted monopole (assuming $w_0 = 1$) and green lines represent the $f_\mrm{NL}$-weighted monopole. Details on the survey assumptions are summarised in Tab.~\ref{tab:surveys}.}
	\label{fig:P0_weighted}
\end{figure*}
%###############################################################

The model to be fitted to the measured, weighted power, also depends on the weights. i.e. we need both the data and model to be sensitive to the same redshifts. We model the theoretical weighted power spectrum multipoles by compressing them into the redshift direction as 
\bea
P_{l,w}(k)\equiv \frac{1}{N_i} \int d\mathcal{W}(z) w_{l,i}(z) P_l(k,z) \label{eq:P0zweighted}
\eea
with
\bea
w_\mrm{unweighted} &=& 1 \\
w_{0,f_\mrm{NL}} &=& \left( b  +\frac{1}{3} f \right) (b-p) D(z) \\
w_{2,f_\mrm{NL}}&=&\frac{4}{3} f (b-p) D(z) 
\eea
with the normalisation $N_i$ given by the equation~(\ref{eq:norm}). %\tcr{Comment on estimators!}

In general, the theoretical power spectrum includes a convolution with the survey window function. However, considering the galaxy power spectrum as an evolving quantity requires a redefinition of the survey window function (for details see Ruggeri et al., in prep.).

%We simulate the redshift weighted power spectrum using Lognormal mock catalogues \citep{Coles:1991if}.

\section{Testing the redshift weights}
\label{sec:Testingweights}

In order to test the weights, we generate an ensemble of mock catalogues, based on overdensities drawn from a Lognormal distribution  \citep{Coles:1991if}. Lognormal-random fields were used for convenience because they approximate the present-day non-linear fluctuation field, and they obey the physical limit $\delta>-1$, which means that they can be Poisson sampled to provide a galaxy distribution with shot-noise and sample variance matching those expected. Although both the assumptions of a Lognormal overdensity field and Poisson-sampled galaxies are crude approximations, they are fit for our purpose of testing the weights.

\begin{table*}
	\centering
	\caption{ We are modelling the eBOSS quasar sample and the DESI ELG sample with the number of galaxies given in Tabel 2 of \citet{2016MNRAS.457.2377Z} and Table 2.3 in  \citet{2016arXiv161100036D} respectively. We are not considering the complete surveys but rather select specific samples to highlight the range of results that can be expected for different survey specification.}
	\label{tab:example_table}
	\begin{tabular}{lcccr} % four columns, alignment for each
		\hline
		survey & tracer&redshift range & sky coverage  & bias model  \\
		\hline
		eBOSS &  Quasars&$0.6<z<2.2$ &  7,500  $\mrm{deg}^2$ &  $b(z)=0.53+0.29(1+z)^2$\\
		DESI & ELGs&$0.6<z<1.8$ & 14,000 $\mrm{deg}^2$ & $b(z)=0.84/D(z)$\\
		\hline
	\end{tabular}
	\label{tab:surveys}
\end{table*}
We generate 10,000 mock catalogues in redshift shells of $\delta z=0.025$ with the number densities, redshift range, sky coverage and bias model as expected for the eBOSS quasar sample and DESI ELGs. A summary of the survey specifications can be found in Table~\ref{tab:surveys}. We assume a box size of  $L=V^{1/3}$ with the volume referring to the shell of a given survey calculated as
\bea
V(z) = \frac{4 \pi}{3} f_\mrm{sky} \left(\chi(z_\mrm{max})^3 - \chi(z_\mrm{min})^3 \right)
\eea
with the sky coverage fraction $f_\mrm{sky}$ and the comoving distance $\chi$.
Within each redshift shell we assume no density gradient, simplifying our analysis to avoid a detailed modelling of survey window function. The simulations assume a flat $\Lambda$CDM cosmology with $\Omega_m=0.3$, $\Omega_b=0.045$, $h=0.7$, $n_s=1.0$, $\sigma_8=0.8$ and $f_\mrm{NL}=0$ as our fiducial cosmology.
We compute the spherically averaged power-spectrum monopole in 23 bins of width $\Delta k=0.005 \ h/\mrm{Mpc}$ from $0.005 \ h/\mrm{Mpc}<k<0.12 \ h/\mrm{Mpc}$ using
\bea
P_0(k,z)=\frac{3}{2} \sum |\tilde{\delta}(k)|^2 \mathcal{L}_0(\mu(k))
\eea
where $\mathcal{L}_0(\mu)$ is the 0th order Legendre polynomial and $|\tilde{\delta}(k)|^2$ is the squared modulus of the Fourier transform of the overdensity $\delta(r)$ at position $r$ and the sum is over all wavevectors in the range $|k|\pm \Delta k/2$ \citep[see  e.g., ][]{2016MNRAS.463.2708P}. For each mock we then calculate the weighted and unweighted power spectra via equation~(\ref{eq:P0zweighted}).

In the following analysis we only consider constraints from the monopole as a proof of concept and do not consider constraints from the quadrupole since most of the information on $f_\mrm{NL}$ is contained in the monopole \citep{2013MNRAS.428.1116R}. For each mock the weighted power spectrum is then calculated using equation~(\ref{eq:P0zweighted}). We calculate the covariance matrix as 
\bea
C_{ij}=\frac{1}{N_m-1} \sum_{n=1}^{N_m} \left[ d_n(k_i) - \bar{d}(k_i) \right]\left[ d_n(k_j) - \bar{d}(k_j) \right]
\eea
where $N_m$ is the total number of mocks, $d_n(k)$ is the power spectrum monopole from the $n$th mock. 

Fig.~\ref{fig:P0_weighted} shows the $f_\mrm{NL}$-weighted and 'unweighted' power spectrum monopole for eBOSS [left panel] and DESI [right panel]. The effect of the $f_\mrm{NL}$-weights is greater for eBOSS due to the adaption of a bias model that evolves more strongly with redshift as well as due to the larger redshift range of the survey. Note, that the normalisation factor for both surveys is different. % \tcr{More discussion on this plot?}  

\begin{figure*}
	\includegraphics[width=0.49\textwidth]{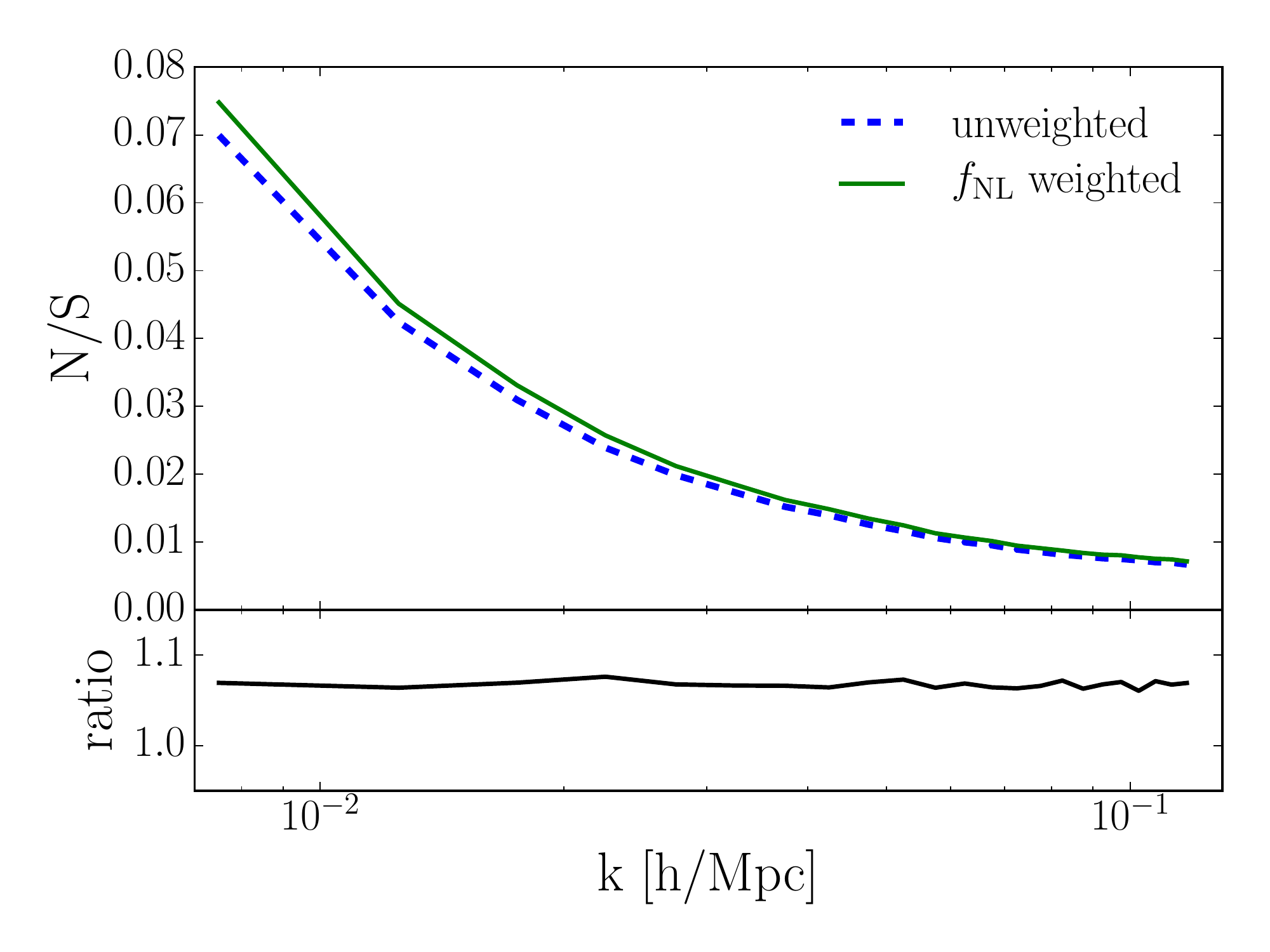}
	\includegraphics[width=0.49\textwidth]{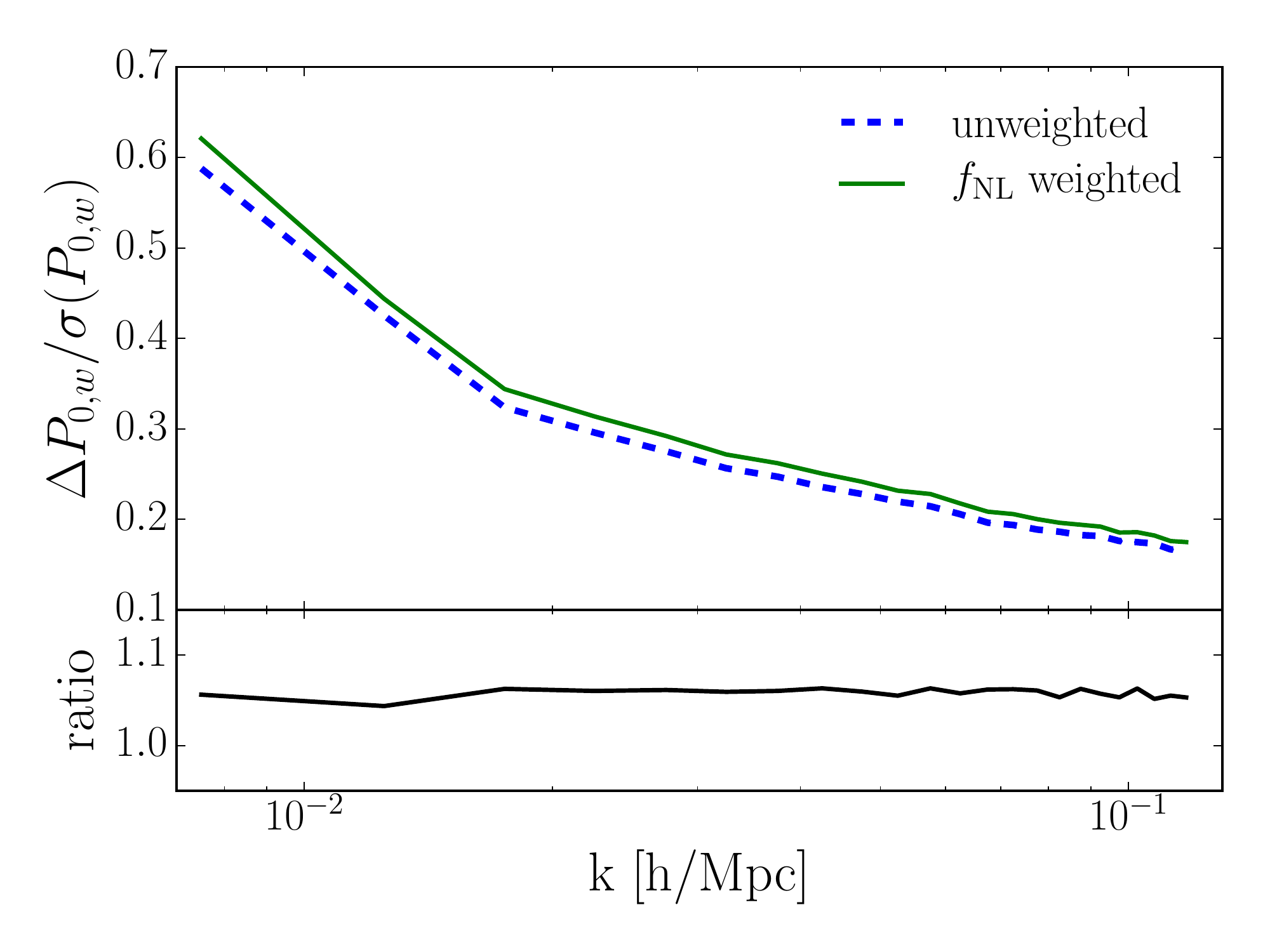}
  	\caption{Noise-to-signal of the weighted power spectrum monopole [left panel] and the difference of the weighted power spectrum assuming  $f_\mrm{NL}=10$ and $f_\mrm{NL}=0$ over the noise, $\Delta P_{0,w}/\sigma(P_{0,w}) =\left( P_{0,w}(f_\mrm{NL}=10) - P_{0,w}(f_\mrm{NL}=0)\right)/\sigma(P_{0,X})$, [right panel]. Blue dashed lines refer to the 'unweighted' power spectrum monopole and green lines to the $f_\mrm{NL}$-weighted monopole. The lower panel shows the ratio between the 'unweighted' and $f_\mrm{NL}$-weighted case. Even though the unweighted monopole has a higher level of noise, the sensitivity to $f_\mrm{NL}$ is higher for the $f_\mrm{NL}$-weighted monopole and with that the capacity to constrain to $f_\mrm{NL}$.}
		\label{fig:NS_weighted}
\end{figure*}

 The redshift weighting scheme takes the redshift evolution of the underlying theory into account, potentially shifting the weights towards regions with higher noise in the clustering signal. Therefore, applying redshift weights does not automatically lead to higher signal to noise in the power spectrum itself. Instead, redshift weighting leads to the observable that can constrain the underlying theory the most. In the case of local non-Gaussianity, more weight is given to galaxies at higher redshifts despite the larger statistical uncertainty at these redshifts, because the effect of $f_\mrm{NL}$ on the powers spectrum is greater at higher redshifts. Fig.~\ref{fig:NS_weighted} depicts the noise-to-signal as a function of scale for the $f_\mrm{NL}$-weighted and 'unweighted' power spectrum monopole for DESI [left panel], as well as the difference of the redshift weighted power spectrum for $f_\mrm{NL}=10$ and $f_\mrm{NL}=0$ over the noise [right panel]. Even though the N/S is larger for the $f_\mrm{NL}$-weighted power spectrum, it has a greater capability to constrain $f_\mrm{NL}$ than the FKP-weighted power spectrum because it is more sensitive to the $f_\mrm{NL}$.

%##########################FIG3###################################
%################################################################

To get an estimate on the error achievable on a measurement of $f_\mrm{NL}$ from both surveys, we calculate the $\chi^2$ surface

%To get an estimate on the uncertainty of $f_\mrm{NL}$, 
%we minimise the $\chi^2$ goodness-of-fit
\bea
\chi^2=(\vec{m}-\vec{d})^T C^{-1} (\vec{m} -\vec{d})
\eea
where $\vec{d}$ is the data vector calculated from the mocks and $\vec{m}$ is the model vector. For both surveys we can then calculate the expected likelihood. For eBOSS we find an uncertainty on $f_\mrm{NL}$ of $\sigma( f_\mrm{NL})=21.63$ at 68\% C.L. for the FKP-weighted case and $\sigma( f_\mrm{NL})=16.66$ for the $f_\mrm{NL}$-weighting scheme, an improvement of 30\%. The improvement for DESI is slightly lower at around 6\%. Our analysis currently uses a scale-dependent FKP weight (i.e. P is allowed to vary with $k$ in the weights). If the FKP weight were fixed, as is often assumed when analysing data for simplicity, then we would expect less good constraints on $f_\mrm{NL}$ because of increased cosmic variance and/or shot noise. We would also have a different fractional improvement from the redshift weights, with the improvement increasing if the FKP weights are fixed for $P(k)$ with $k$ on larger scales: those where the $f_\mrm{NL}$ signal is stronger and the redshift-weights more effective. For example if the FKP weight is fixed at $k_0=0.0475$ the improvement increases to 42\% for eBOSS.

The Fisher matrix forecasts for eBOSS quasars are $\sigma( f_\mrm{NL})=15.74$ \citep{2016MNRAS.457.2377Z} with fixing the bias. The redshift weighting technique yields results closer to the predicted uncertainty compared to the unweighted analysis. We do not quite reach the Fisher forecast accuracy because we only consider the monopole and assume a slightly smaller $k$-range. The Fisher forecasts for DESI are $\sigma( f_\mrm{NL})=3.8$ \citep{2014JCAP...05..023F}, yet these constraints are for the full DESI survey and not just the ELG sample.

 The difference between the improvement for eBOSS and DESI from adding the new weights is driven by the range of bias assumed across the sample under consideration, and so will not be fully known for DESI until the survey starts. Even so, this analysis is a proof of principle that the fNL-redshift weighting can lead to stronger constraints on fNL than a simple FKP-weighted power spectrum.

%This analysis is a proof of principle that the $f_\mrm{NL}$-redshift weighting can lead to stronger constraints on $f_\mrm{NL}$ than a simple FKP-weighted power spectrum. We can also access the relative improvement for different upcoming surveys. In principle, the improvement from the $f_\mrm{NL}$-weighting increases with the redshift range of the survey but also depends on the tracer of the sample, with the improvement being stronger where the range of bias across a sample is larger.

%#########################FIG4##################################
\begin{figure*}
	\label{fig:Likelihood}
	\includegraphics[width=0.49\textwidth]{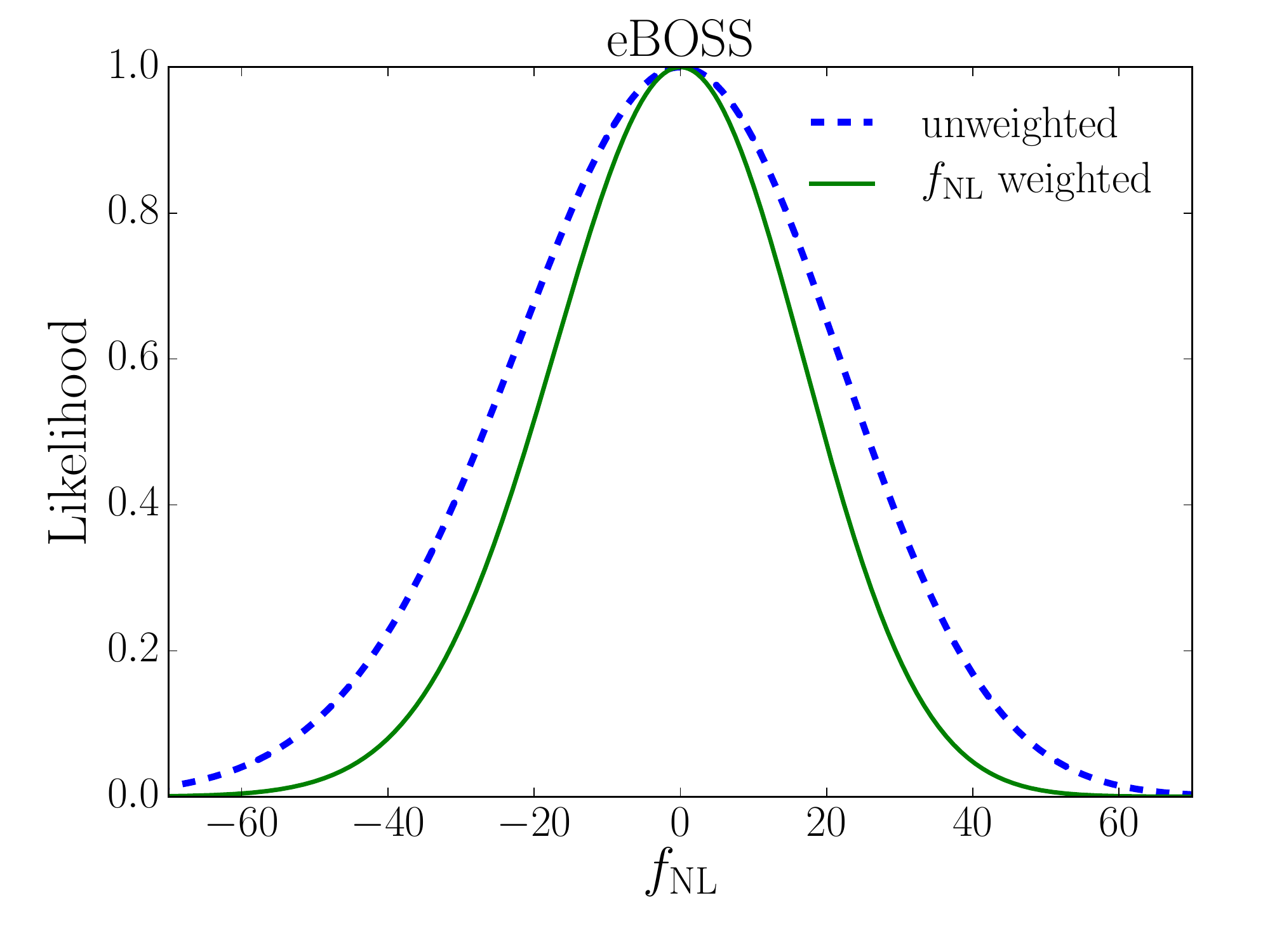}
	\includegraphics[width=0.49\textwidth]{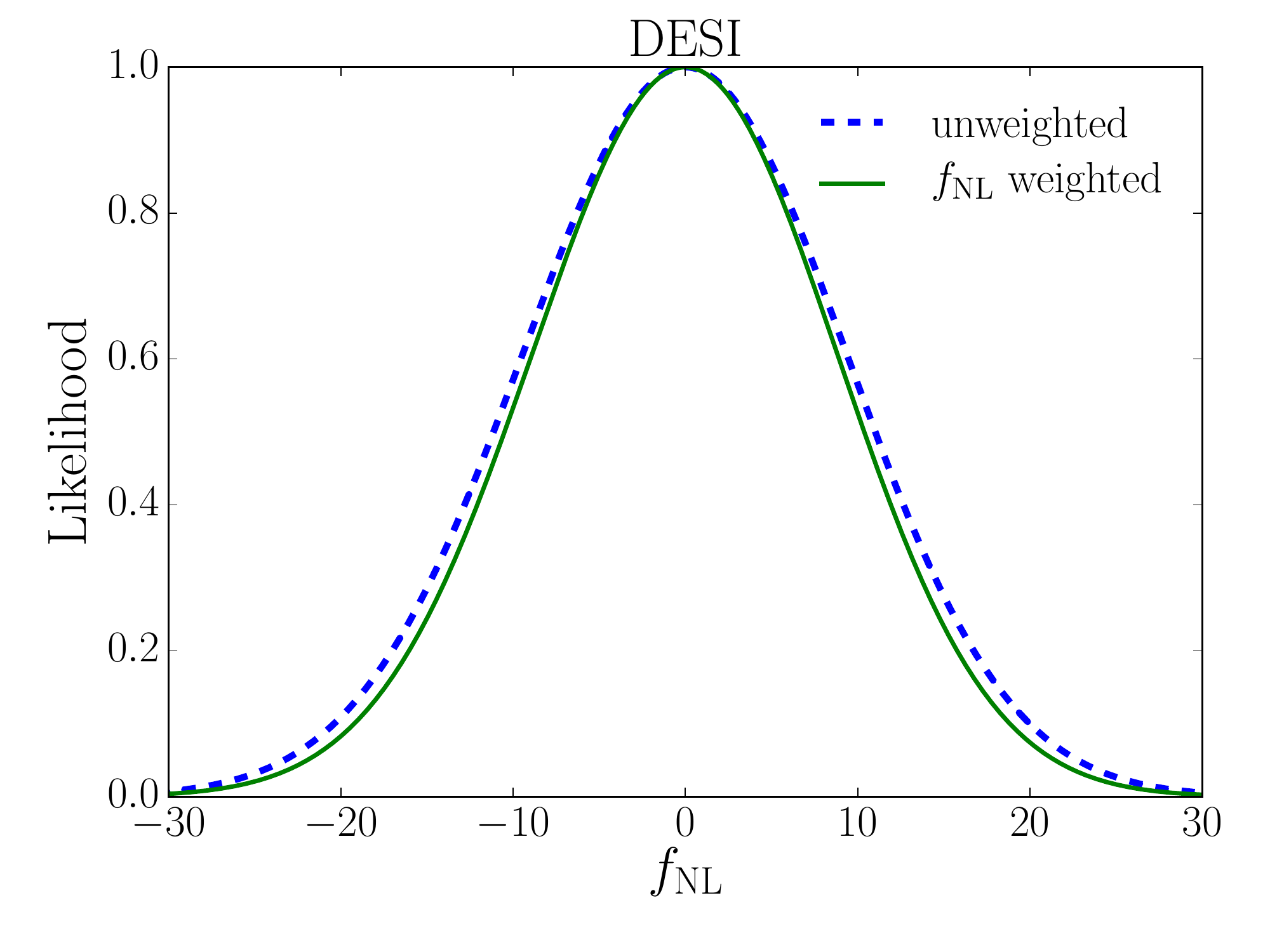}
  	\caption{ Projected likelihood for $f_\mrm{NL}$ measurements from eBOSS quasars [left panel] and DESI ELGs [right panel] for an 'unweighted' (green lines) and $f_\mrm{NL}$-weighted (blue dashed lines) power spectrum monopole. The redshift weighting technique can improve the constraints on $f_\mrm{NL}$ by 30\% for eBOSS quasars and  6\% for DESI ELGs. The improvement is larger for eBOSS due to the strong redshift evolution of the assumed bias model as well as a larger redshift range.}
\end{figure*}
%#############################################################

%%%%%%%%%%%%%%%%%%% CONCLUSION %%%%%%%%%%%%%%%%%%%
\section{Conclusions}
\label{sec:conclusion}

The scale dependent bias is a strong probe of non-Gaussianity and upcoming LSS surveys will put tight constraints on the amplitude of the primordial fluctuations, $f_\mrm{NL}$. Since these surveys cover large redshift ranges, redshift weighting, a new analysing technique that does not rely on binning in redshift slices, provides a promising way of fully exploiting LSS information. Redshift weighting is in particular important for studying the primordial universe since non-Gausssianity alters large scales more strongly than small scales. Not splitting the sample into small redshift slices therefore increases the effective number of relevant scales modes that are included in any analysis. Indeed, applying redshift weights will be crucial to reach the accuracy predicted by Fisher forecasts, which implicitly assume that all of the information is extracted, in effect assuming optimised weights are used.

%In this paper we have derived the optimal weights to measure $f_\mrm{NL}$ using the redshift weighting formalism to perfectly balance cosmic variance, shot noise and the redshift evolution of the scale dependent halo bias induced by non-Gaussianity. The weights depend on the assumed tracer of the galaxy sample as well as the assumed bias model, generally increasing with redshift, therefore weighting galaxies at high redshift stronger than at low redshifts.
The optimal weights we have derived to measure $f_\mrm{NL}$ balance sample variance, shot noise and the redshift evolution of the scale dependent halo bias induced by non-Gaussianity. The weights depend on the properties of the galaxy sample through the evolving bias of the sample. As the bias is generally increasing with redshift, we end up weighting galaxies at high redshift more strongly than at low redshifts, even if the signal-to-noise of the clustering signal is weaker.

We assessed the potential of the $f_\mrm{NL}$ weights using mock catalogs generated though a Lognormal code simulating the upcoming eBOSS and DESI surveys. We find that the uncertainty on $f_\mrm{NL}$ is minimised when applying the $f_\mrm{NL}$ redshift weights, yielding an improvement of 6\% up to 30\% for DESI and eBOSS respectively compared to analysing an FKP-weighted power spectrum. %The improvement increases with a larger redshift range of the survey and with a stronger redshift dependent bias model.

There are a few caveats to our analysis: First, the redshift weights to optimally measure local non-Gaussianity depend strongly on the assumed galaxy bias.
If the fiducial bias model is inaccurate, then the weights will not be optimal and lead to looser constraints on $f_\mrm{NL}$ then expected. However, the redshift weighted power spectrum will still be unbiased. Second, for tracers with no strongly evolving bias the underlying theory is only mildly redshift dependent limiting the overall improvement of the redshift weighting technique. In general, the improvement from the $f_\mrm{NL}$-weighting increases with the redshift range of the survey but also depends on the tracer of the sample, with the improvement being stronger where the range of bias across a sample is larger.

We have discussed the potential of redshift weighting to constrain $f_\mrm{NL}$ for eBOSS and DESI, but there are also other future surveys for which this technique is highly relevant, for instance the Euclid mission \citep{2013LRR....16....6A} and SPHEREx \citep{2014arXiv1412.4872D}. Euclid is a space based, spectroscopic survey of H$\alpha$-selected emission line galaxies with galaxies in redshift range range $0.7<z<2.0$, expected to constrain local non-Gaussianity in addition to BAO and RSD measurements. SPHEREx is a all-sky spectroscopic satellite survey covering a very wide redshift range that was particularly designed to measure non-Gaussianity. It has an evolving redshift accuracy up to $\sigma_z/(1+z)<0.2$ with low redshifts being more accurately measured than high redshifts. However, we expect the lower redshift accuracy not to be problematic when applying the redshift weights as long as the uncertainty in redshift is taking into account as an additional contribution to the covariance when calculating the weights. 

In this work, we have only considered weights optimised to measure non-Gaussianity in the local framework. We leave the study of weights for more complex models, such as equilateral or orthogonal shapes, models with non-zero running of $f_\mrm{NL}$, or shapes with specific angle dependency for future work. Additionally, a natural extension of this work is to apply the redshift weighting technique to multiple tracer samples, therefore combining the optimal redshift weights with weights designed to optimally exploit the additional information through multi tracer methods \citep{2016MNRAS.463.2708P,2011PhRvD..84h3509H}.

\section*{Acknowledgements}
We thank Alkistis Pourtsidou and Fangzhou Zhu for insightful discussions and comments. 
EM, RR and WJP acknowledge support from the European Research Council through the Darksurvey grant 614030. WJP also acknowledges support from the UK Science and Technology Facilities Council grant ST/N000668/1 and the UK Space Agency grant ST/N00180X/1.

%%%%%%%%%%%%%%%%%%%%%%%%%%%%%%%%%%%%%%%%%%%%%%%%%%



\bibliographystyle{mnras}
\bibliography{weights.bib} % if your bibtex file is called example.bib


%%%%%%%%%%%%%%%%%%%%%%%%%%%%%%%%%%%%%%%%%%%%%%%%%%

%%%%%%%%%%%%%%%%%% APPENDICES %%%%%%%%%%%%%%%%%%%%%
%
%\appendix

% Don't change these lines
\bsp	% typesetting comment
\label{lastpage}
\end{document}